\newcommand{\ud}[1]{{#1^{\dagger}}}
\newcommand{\bra}[1]{\left\langle #1\right|}
\newcommand{\ket}[1]{\left| #1\right\rangle}

\documentclass[
    ,final            
  ]
  {aipproc}

\usepackage{amsmath}

\layoutstyle{6x9}

\begin{document}

\title{Optical Spectra of the Jaynes-Cummings Ladder}


\author{Fabrice P. Laussy}{
  address={School of Physics and Astronomy, University of Southampton, United Kingdom}
}

\author{Elena del Valle}{
  address={Departamento de Fisica Teorica de la Materia Condensada (C--V), Universidad Autonoma de Madrid, Spain}
}

\begin{abstract}
  We explore how the Jaynes-Cummings ladder transpires in the emitted
  spectra of a two-level system in strong coupling with a single mode
  of light. We focus on the case of very strong coupling, that would
  be achieved with systems of exceedingly good quality (very long
  lifetimes for both the emitter and the cavity). We consider the
  incoherent regime of excitation, that is realized with
  semiconductors quantum dots in microcavities, and discuss how
  reasonable is the understanding of the systems in terms of
  transitions between dressed states of the Jaynes-Cummings
  Hamiltonian.
\end{abstract}

\maketitle


\section{Introduction}

The crowning achievement of cavity Quantum Electrodynamics is the
so-called \emph{Strong Coupling} (SC), a regime dominated by quantum
interactions between light (photons) and matter (an atom, or its
semiconductor realization, a quantum dot (QD), sometimes called an
artificial atom)~\cite{kavokin_book07a}. In this regime, both the atom
and the photon loose their individual identity, and vanish to give
rise to new particles---sometimes called \emph{polaritons}---with new
properties. In this text, we shall be concerned with their optical
spectral properties.

Strong light-matter coupling originated with atomic
cavities~\cite{raizen89a,mabuchi02a}, and was later reproduced (among
other systems) with semiconductor cavities~\cite{weisbuch92a}, more
promising for future technological applications (see for
instance~\cite{laucht08a}). It is only recently, however (c.~2004),
that this regime was reached for the zero-dimensional semiconductor
case~\cite{reithmaier04a,yoshie04a,peter05a} and the number of reports
has not been overwhelming ever
since~\cite{hennessy07a,press07a,nomura08a,laucht08a}. In these
systems, the QD excitation---the exciton---strongly couples to the
single mode of a microcavity (realized as a pillar, a photonic crystal
or other variants). As far as cavity QED is concerned, these systems
are in principle superior to their two-dimensional counterpart (where
SC is routinely achieved), because only a few excitations enter the
problem in an environment with much reduced degrees of freedom, as
opposed to planar polaritons whose most adequate description is in
terms of continuous fields.\footnote{Here we must outline that we mean
  ``quantum'' in the sense of quantization of the fields, and thus
  breaking the classical picture. 2D~polaritons have demonstrated
  stunning properties rooted in quantum physics, such as Bose-Einstein
  condensation or superfluid
  motion~\cite{kasprzak06a,amo09a}. However, those are manifestations
  of macroscopic coherence where large numbers of microscopic
  particles exhibit the behaviour of a continuous field (classical or
  not).}

The landmark of SC is the Rabi doublet, where two modes (light and
matter) at resonance, do not superimpose but split, each line
corresponding to one of the polaritons that have overtaken the bare
modes. It was early understood that this splitting is by itself,
however, not a proof of quantization~\cite{zhu90a}.  Parenthetically,
2D~polaritons display most eloquently this splitting (see footnote~1).

Antibunching in the optical emission of a strongly coupled
QD-microcavity system has been
demonstrated~\cite{hennessy07a,press07a}, further supporting
quantization of the fields, but this is not completely conclusive, as
although it proves that the dynamics involves a single quantum of
excitation between two isolated modes (by itself already a
considerable achievement), it does not instruct on the modes
themselves (consider the vacuum Rabi problem of two harmonic
oscillators, that gives the same result).  After all, dimming
classical light until single photons remain, would exhibit
antibunching, but this says nothing about the emitter itself (which is
ultimately quantum anyway; if it's coming from the sun, say, it
originates from the spontaneous emission of an atom, or to much lower
probability, from stimulated emission).

A genuine, or quantum, SC~\cite{khitrova06a}, should culminate with a
direct, explicit demonstration of quantization, with one quantum more
or less changing the behaviour of the system. The most fundamental
model to describe light (bosons)/matter (fermions) interactions is the
celebrated \emph{Jaynes-Cummings} (JC) Hamiltonian~\cite{shore93a},
where such a quantum sensitivity is strongly manifest, and has been
observed more or less directly in various systems~\cite{brune96a,
  meekhof96a}. Recently, direct spectroscopic evidence has been
reported for atoms and superconducting circuits, in elaborate
experiments~\cite{schuster08a, fink08a} that remind the heroic efforts
of Lamb to reveal the splitting of the orbitals of hydrogen.  Even
more recently, very clear transitions from up to the fifth step of the
ladder have been unambiguously observed in circuit QED in very
strong-coupling, with the Rabi splitting more than~260 times the
vacuum linewidth~\cite{bishop09a}!

With semiconductor QDs, there has been so far, to the best of our
knowledge, no explicit demonstration of JC nonlinearities. In a
previous work~\cite{osjcl_delvalle08b}, we analyzed the peculiarities
of these systems both with respect to their particular physics
(involving a steady state under incoherent excitation rather than
spontaneous emission or coherent excitation~\cite{laussy08a}) and
their parameters. We concluded that even present-day structures could
evidence anharmonicity of the quantized levels with particular pumping
schemes and a careful analysis of the data. With much better
structures but still conceivably realizable in the near future, strong
qualitative signatures emerge and there is no need to go further than
a simple observation of anharmonically spaced multiplets. Here we take
the further step to go towards unrealistically good (as of today!)
structures, with negligible exciton decay and quality factors orders
of magnitude higher than state of the art system. In this regime of
very strong coupling, one expects a priori Lorentzian emission of the
dressed states~\cite{laussy05c}. We show with an exact quantum-optical
computation of the spectra, the value and limitations of this
approximation.  Our results below can be seen as the ideal quantum
limit of the Jaynes-Cummings model of light and matter, where the
quantization of the fields appears in its full bloom.

\section{The Jaynes-Cummings physics}

\begin{figure}
  \centering
  \includegraphics[width=.85\linewidth]{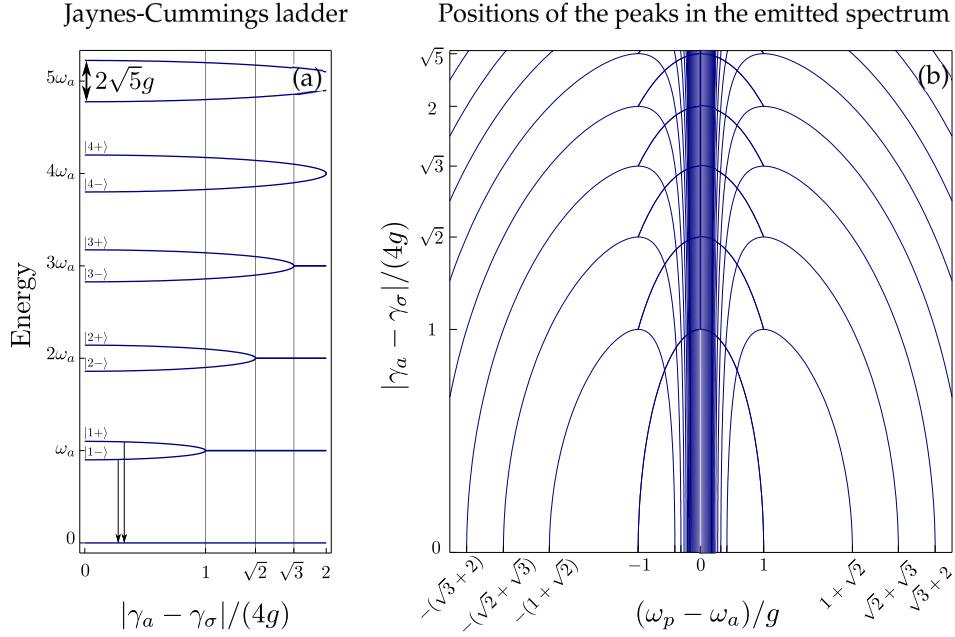}
  \caption{The Jaynes-Cummings ladder (a) and the positions of peaks
    in its emitted spectrum (b), obtained from the difference in
    energy of transitions between any two branches of two consecutives
    ``steps'' (or manifolds) of the ladder. Without dissipation, the
    splitting of the $n$th step is~$2\sqrt{n}g$ and transitions
    produce an infinite sequence of peaks
    at~$\pm\sqrt{n}\pm\sqrt{n+1}$, an infinitely countable number of
    them piling up towards 0 from above and below. With
    increasing~$|\gamma_a-\gamma_\sigma|$, the steps go one after the
    other into weak-coupling, producing a complex diagram of
    branch-coupling in the resonances of the emitted spectrum (b).}
  \label{fig:WedJan21174936GMT2009}
\end{figure}

The Jaynes-Cummings Hamiltonian is a textbook model, that admits
essentially analytical solutions. It reads:
\begin{equation}
  \label{eq:TueJul15133406CEST2008}
  H=\omega_a\ud{a}a+\omega_\sigma\ud{\sigma}\sigma+g(\ud{a}\sigma+a\ud{\sigma})\,.
\end{equation}
Here, $\omega_{a,\sigma}$ are the free energies for the modes $a$
(boson) and~$\sigma$ (fermion) and~$g$ is their coupling strength.  To
describe realistically an experiment, one needs at least to include
dissipation. Excitation is then typically assumed as an initial state,
or with a coherent (Hamiltonian) pumping. To address the semiconductor
case, we considered incoherent excitation~\cite{osjcl_delvalle08b},
i.e., a rate~$P_\sigma$ of QD excitation.\footnote{We also considered
  a rate~$P_a$ of incoherent \emph{photon} excitation, due to other
  dots or any other source populating the cavity. For more dissipative
  cases, this term bears a crucial importance on the spectral
  shapes. However in the very strong coupling, its role is of a less
  striking character and we therefore ignore it in the present
  discussion.} The equation of motion for the density matrix~$\rho$
then
reads~$\partial_t\rho=i[H,\rho]+\sum_{c=a,\sigma}\gamma_c(2c\rho\ud{c}-\ud{c}c\rho-\rho\ud{c}c)+P_\sigma(2\ud{\sigma}\rho\sigma-\sigma\ud{\sigma}\rho-\rho\sigma\ud{\sigma})$.
So far we have not been able to provide an analytical solution for the
steady state density matrix of this equation, but in the tradition of
the JC, many exact results can nevertheless be extracted, for instance
the eigenvalues of~$H$ in presence of dissipation, in
Fig.~\ref{fig:WedJan21174936GMT2009}(a). These give the energies of
the states that are renormalized by the light-matter interaction,
Eq.~(\ref{eq:TueJul15133406CEST2008}). For a given number~$n$ of
excitations, there are two new eigenstates~$\ket{n+}$ and~$\ket{n-}$
that take over the bare modes~$\ket{n~\mathrm{photon(s)},
  0~\mathrm{exciton}}$ and $\ket{n-1~\mathrm{photon(s)},
  1~\mathrm{exciton}}$. The polaritons are, in their canonical sense,
the states~$\ket{1,\pm}=(\ket{10}\pm\ket{01})/\sqrt2$. For each~$n$,
the two new states are splitted by $2\sqrt{n}g$. This is the main
manifestation of quantization in the JC Hamiltonian: the difference
from~$\sqrt{n}$ to~$\sqrt{n+1}$, when~$n$ is small, can be detected in
a careful experiment. This structure of renormalized states is called
the \emph{Jaynes-Cummings ladder}. The transitions between its steps
account for the lines that are observed in an optical luminescence
spectrum. This generates the structure shown on
Fig.~\ref{fig:WedJan21174936GMT2009}(b).\footnote{This is only the
  imaginary part. The real part, that corresponds to broadening of the
  transitions, is given in
  Ref.~\cite{osjcl_delvalle08b}.} 

The transitions between the same kind of dressed states (``same-states
transitions'') of two adjacent steps (or \emph{manifolds},
mathematically speaking), e.g., from~$\ket{n+}$ to~$\ket{n-1,+}$, emit
at the energy~$\sqrt{n+1}+\sqrt{n}$ (in units of~$g$), while the
transitions between different kind of dressed states
(``different-states transitions'')e.g., from~$\ket{n+}$
to~$\ket{n-1,-}$ emit at~$\sqrt{n+1}-\sqrt{n}$. Different-states
transitions emit beyond the Rabi doublet while the same-states ones
pack-up in between.

To obtain the exact emission spectra of this system, one needs to
compute two times correlators~$G^{(1)}_a(t,\tau)=\langle
a(t+\tau)a(\tau)\rangle$ for the cavity emission
and~$\langle\sigma(t+\tau)\sigma(\tau)\rangle$ for the direct exciton
emission (in the steady state, the limit~$t\rightarrow\infty$ is
taken). These two cases correspond to the geometry of detection in an
ideal spectroscopic measurement: the cavity spectrum $S_a(\omega)$
(the $\tau$-Fourier Transform of~$G^{(1)}_a$) corresponds to detection
of the cavity mode---this would be along the cavity axis of a pillar
microcavity, for instance---while~$S_\sigma(\omega)$ (FT
of~$G^{(1)}_\sigma$) corresponds to direct emission of the exciton,
like the top emission of a photonic crystal or the side of a
pillar. We present both cases in the results below.

\section{Results}

\begin{figure}
  \centering
  \includegraphics[width=\linewidth]{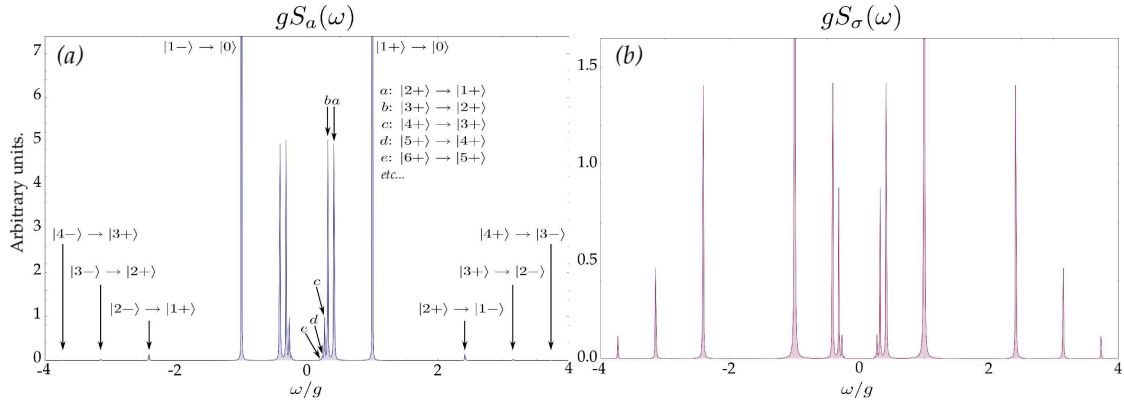}
  \caption{Fine structure of the ``light-matter molecule'': emission
    spectra in the cavity~$(a)$ and direct exciton emission~$(b)$ of
    the strongly-coupled system
    with~$(\gamma_a,\gamma_\sigma)/g=(10^{-3},0)$ at
    $P_\sigma/g=10^{-3}$.}
  \label{fig:MonFeb2072644GMT2009}
\end{figure}

We consider very good cavities with~$g=1$ (defining the unit),
$\gamma_a$ of the order of~$10^{-3}$, $10^{-2}$ and~$\gamma_\sigma=0$
(results are not significantly modified qualitatively for nonzero
values of~$\gamma_\sigma$ of the order of~$\gamma_a$). We consider
small electronic pumping~$P_\sigma$ of the same order
of~$\gamma_a$. At higher pumpings, the system goes into the classical
regime, with lasing, very high populations, and continuous fields
replacing discretization~\cite{osjcl_delvalle08b}. We therefore focus
on small pumping rates, as we are interested in manifestations of
quantization.  Those are neatly displayed, as seen on
Fig~\ref{fig:MonFeb2072644GMT2009}, where very sharp (owing to the
small decay rates) lines reconstruct the transitions of the JC
ladder. In the plot of the cavity emission,
Fig~\ref{fig:MonFeb2072644GMT2009}$(a)$, we have marked each peak with
its corresponding transition between two quantized, dressed states of
the JC hamiltonian. These peaks correspond one-to-one with those of
the exciton emission, that are, however, weighted
differently~\cite{osjcl_delvalle08b}. A simple argument explains the
different weights in the intensity of the lines in the cavity
emission~$I_a$ as compared to their counterpart in the exciton
emission~$I_\sigma$:

\begin{subequations}
  \label{eq:MonFeb2115455GMT2009}
  \begin{align}
    &\mathrm{I}_{a}^{(\pm\rightarrow\mp)}\propto|\bra{n-1,\mp}a\ket{n,\pm}|^2=|\sqrt{n}-\sqrt{n-1}|^2/4\,,\\
    &\mathrm{I}_{a}^{(\pm\rightarrow\pm)}\propto|\bra{n-1,\pm}a\ket{n,\pm}|^2=|\sqrt{n}+\sqrt{n-1}|^2/4\,,
  \end{align}
\end{subequations}
i.e., intensity is enhanced for same-peak but smothered for
different-peak transitions, while on the other hand, the QD emission
is level regardless of the configuration:
\begin{subequations}
  \label{eq:ThuNov27133856GMT2008}
  \begin{align}
    &\mathrm{I}_{\sigma}^{(\pm\rightarrow\mp)}\propto|\bra{n-1,\mp}\sigma\ket{n,\pm}|^2=1/4\,,\\
    &\mathrm{I}_{\sigma}^{(\pm\rightarrow\pm)}\propto|\bra{n-1,\pm}\sigma\ket{n,\pm}|^2=1/4\,.
  \end{align}
\end{subequations}

\begin{figure}
  \centering
  \includegraphics[width=\linewidth]{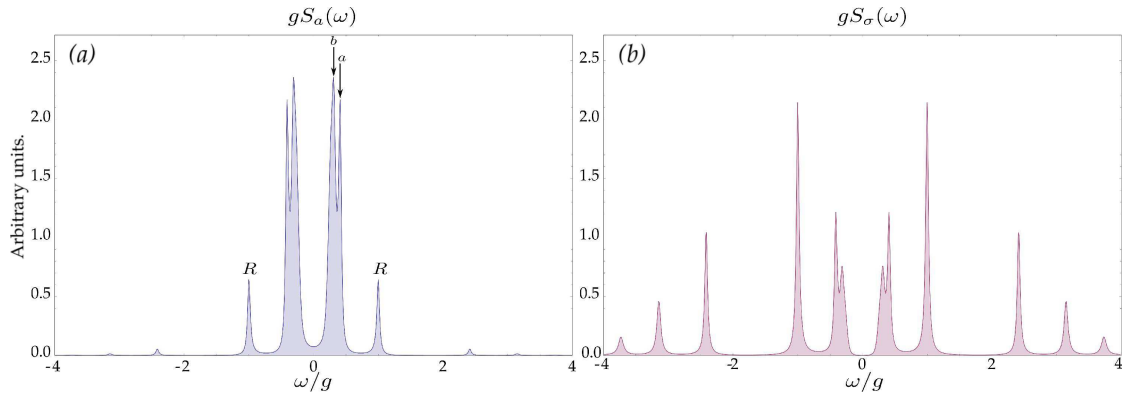}
  \caption{Same as Fig.~\ref{fig:MonFeb2072644GMT2009} but now
    with~$\gamma_a/g=10^{-2}$. Less peaks are resolved because of
    broadening but nonlinear peaks ($a$, $b$) are neatly
    observable. In fact, now inner nonlinear peaks dominate in the
    cavity emission (the linear Rabi peaks are denoted~$R$). In the
    exciton direct emission, the Rabi doublet remains the strongest.}
  \label{fig:MonFeb2075641GMT2009}
\end{figure}

In both cases, the Rabi doublet dominates strongly over the other
peaks. In Fig.~\ref{fig:MonFeb2072644GMT2009}$(a)$, for instance, the
peaks at~$\pm1$ extend for about 9~times higher than is shown, and
already the outer transitions are barely noticeable. This is because
the pumping is small and so also the probability of having more than
one photon in the cavity (it is in this configuration of about~10\% to
have~$2$ photons, see Fig.~\ref{fig:MonFeb2085118GMT2009}). One could
spectrally resolve the window~$[-g/2,g/2]$ over a long integration
time and obtain the multiplet structure of nonlinear inner peaks, with
spacings~$\{\sqrt{n+1}-\sqrt{n},\, n>1\}$ (in units of~$g$), observing
direct manifestation of single photons renormalizing the quantum
field. Or one could increase pumping (as we do later) or use a cavity
with smaller lifetime. In this case, less peaks of the JC transitions
are observable because of broadenings mixing them together, dephasing
and, again, reduced probabilities for the excited states, but the
balance between them is better. In
Fig.~\ref{fig:MonFeb2075641GMT2009}, where~$\gamma_a$ is now~$g/100$,
the Rabi doublet (marked~$R$) is dominated by the nonlinear inner
peaks in the cavity emission, and a large sequence of peaks is
resolved in the exciton emission.

\begin{figure}
  \centering
  \includegraphics[width=\linewidth]{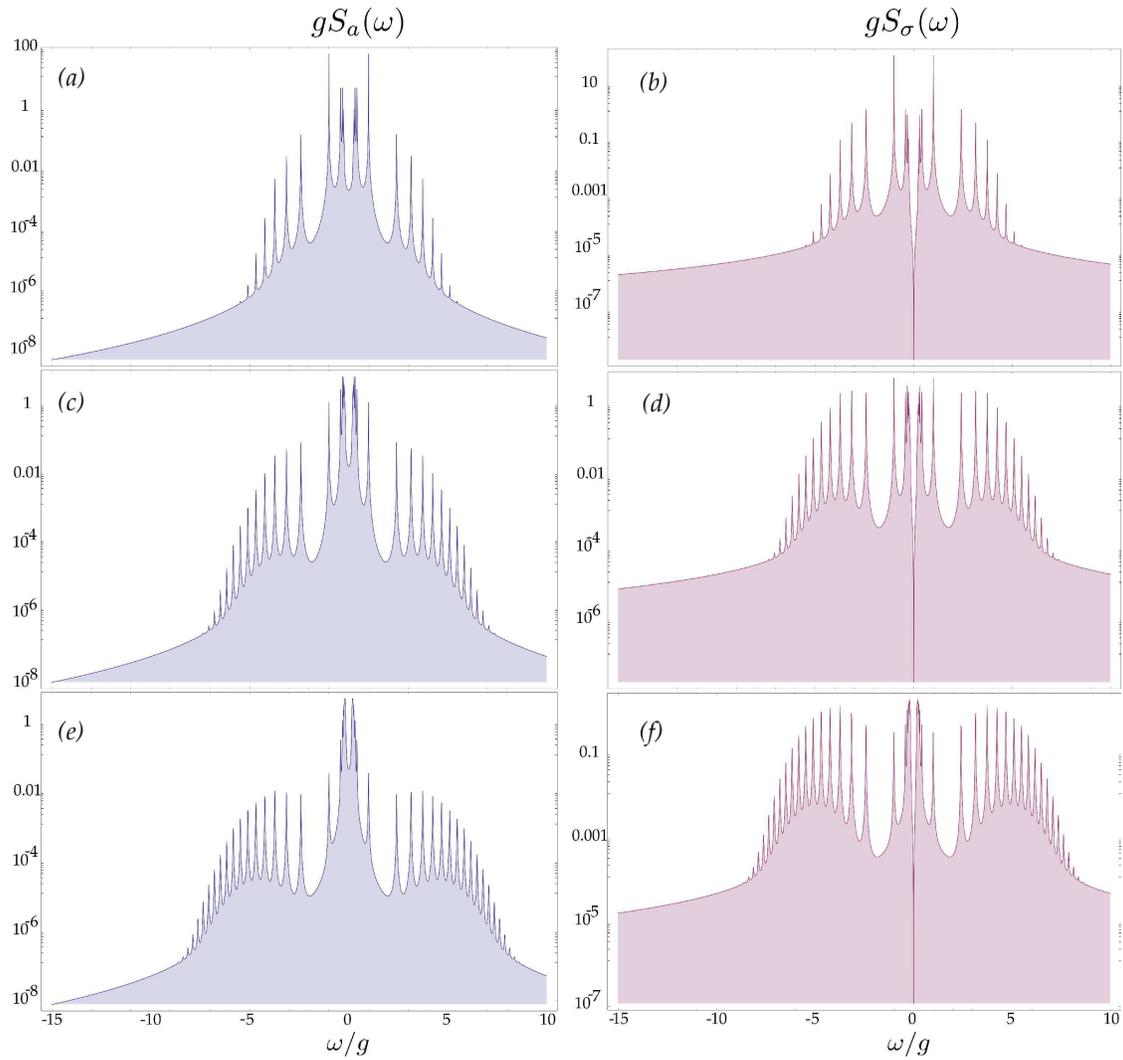}
  \caption{Spectra of emission in log-scales as a function of
    pumping~$P_\sigma/g$, for $10^{-3}$ (upper row), $5\times10^{-3}$
    (middle) and~$10^{-2}$ (lower row).}
  \label{fig:MonFeb2082806GMT2009}
\end{figure}

\begin{figure}
  \centering
  \includegraphics[width=.5\linewidth]{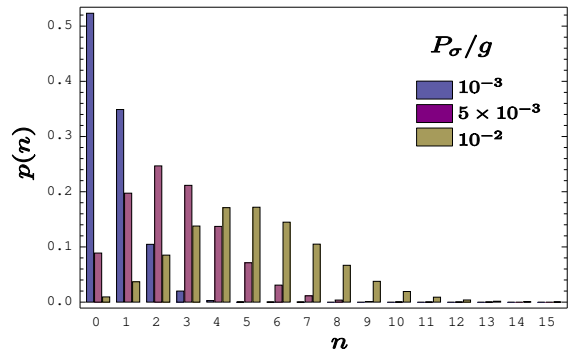}
  \caption{Probability~$p(n)$ of having~$n$ photon(s) in the cavity,
    for the three cases shown on
    Fig.~\ref{fig:MonFeb2082806GMT2009}. Quite independently of the
    distribution of photon numbers in the cavity, field-quantization
    is obvious.}
  \label{fig:MonFeb2085118GMT2009}
\end{figure}

Going back to the case of Fig.~\ref{fig:MonFeb2072644GMT2009}, but
increasing pumping, we observe the effect of climbing higher the
Jaynes-Cummings ladder. Results are shown in
Fig.~\ref{fig:MonFeb2082806GMT2009} in logarithmic scale, so that
small features are magnified. First row is
Fig.~\ref{fig:MonFeb2072644GMT2009} again but in log-scale, so that
the effect of this mathematical magnifying glass can be
appreciated. Also, we plot over the wider range~$[-15g,10g]$. Note how
the fourth outer peak, that was not visible on the linear scale, is
now comfortably revealed with another three peaks at still higher
energies. As pumping is increased, we observe that the strong linear
Rabi doublet is receding behind nonlinear features, with more
manifolds indeed being probed, with their corresponding transitions
clearly observed (one can track up to the 19th manifold in the last
row). This demonstrates obvious quantization in a system with a large
number of photons. The distribution of photons in these three cases is
given in Fig.~\ref{fig:MonFeb2085118GMT2009}, going from a
thermal-like, mostly dominated by vacuum, distribution, to
coherent-like, peaked distribution stabilizing a large number of
particles in the system.  At the same time, note the cumulative effect
of all the side peaks from the higher manifolds excitations, absorbing
all quantum transitions into a background that is building up
shoulders, with the overall structure of a triplet. This is the
mechanism through which the system bridges from a quantum to a
classical system. The new spectra are reminiscent of the Mollow
triplet of resonance fluorescence~\cite{mollow69a}. These are obtained
both in the cavity and the exciton emission, but much more so in the
latter. With more realistic parameters, it is indeed only visible in
the exciton emission.

In the very strong coupling limit, one could content with a basic
understanding of the transitions mechanisms (given by
Eqs.~(\ref{eq:MonFeb2115455GMT2009}-\ref{eq:ThuNov27133856GMT2008}))
and the knowledge of the distribution of particles
(cf.~Fig.~\ref{fig:MonFeb2085118GMT2009}), to provide a fairly good
account of the final result~\cite{laussy05c}. This is true for the
most basic understanding of the transitions only, essentially valuable
to identify peaks in terms of transitions in the JC ladder. Some
features of the problem, like interferences, are missing
completely. Observe for instance the sharp dip that is retained in all
the exciton spectra. This is a quantum optical result that can be
recovered only with an exact treatment of the emission. The
qualitative result remains strong enough to support the underlying
physics with little needs of these refinements. In the case of less
ideal cavities, however, the evidence is not so strong for
field-quantization, and both a proper description as well as an
understanding of its specificities is then required to support one's
conclusions.

\section{Conclusion}

We have overviewed the problematic of nonlinearities in optical
spectra as it is posed by cavity QED, where both fields (atomic and
photonic) are quantized. We explored the very strong coupling regime
in a system with very small decay rates. We confirm with an exact
quantum optical treatment that a qualitative picture in terms of
transitions in the Jaynes-Cummings ladder reasonably accounts for the
observations. The successful observations of full quantization of the
light-matter field merely require high experimental sensitivity and
very good samples. In less ideal conditions, a finer analysis is
required, as was discussed elsewhere~\cite{osjcl_delvalle08b}.


\begin{theacknowledgments}
  This work has been initiated in UAM Madrid in the group of
  Prof. Carlos Tejedor. Results have been computed with the Iridis
  cluster facility of the University of Southampton.
\end{theacknowledgments}



\bibliographystyle{aipproc}   

\bibliography{Sci,books,osjcl}

\IfFileExists{\jobname.bbl}{}
 {\typeout{}
  \typeout{******************************************}
  \typeout{** Please run "bibtex \jobname" to optain}
  \typeout{** the bibliography and then re-run LaTeX}
  \typeout{** twice to fix the references!}
  \typeout{******************************************}
  \typeout{}
 }

\end{document}